\documentclass[preprint,preprintnumbers,amsmath,amssymb]{revtex4}

\usepackage{epsfig}
\usepackage{graphicx}% Include figure files
\usepackage{dcolumn}% Align table columns on decimal point
\usepackage{bm}% bold math

\usepackage[normalsize]{caption}%descripcion de figuras

\begin{document}

%................................................
%\begin{figure}[htb]
%\centerline{\includegraphics[scale=0.2]{openwebmail.eps}}
%\caption{IPICyT}
%\end{figure}
%....................................................

\setcounter{MaxMatrixCols}{10}

\title{Genetic Algorithm Optimization of Entanglement}

\author{Jorge C. Navarro-Mu\~noz\footnote{Electronic address:jcarlos@ipicyt.edu.mx},
H.C. Rosu\footnote{Corresponding author. Electronic address:
hcr@ipicyt.edu.mx} and R. L\'opez-Sandoval\footnote{Electronic
address: sandoval@ipicyt.edu.mx}}

\affiliation{$^1$ Potosinian Institute of Science and Technology,\\
Apartado Postal 3-74 Tangamanga, 78231 San Luis Potos\'{\i}, Mexico\\
\rm (Received 19 April 2006; published 7 November 2006)}

%\date{Sept 18, 2006}
\date{Nov. 13, 2006}

%\centerline{ArXiv: q-bio.QM/0409036  [v2]}

\centerline{{\small \tt Phys. Rev. A 74, 052308 (2006)}}

\medskip

\centerline{{\small \tt quant-ph/0604126}}

%,,,,,,,,,,,,,,,,,,,,,,,,,,,,,,,,,,,,,,,,,,,,,,,,,,,,,,,,,,,,,,,,,,,,,,,,,,,,,
\begin{abstract}
We present an application of a genetic algorithmic computational
method to the optimization of the concurrence measure of
entanglement for the cases of one dimensional chains, as well as
square and triangular lattices in a simple tight-binding approach in
which the hopping of electrons is much stronger than the phonon
dissipation.
%The results could be relevant for lattices and small
%clusters (nanoparticles) of magnetic oxides and fluorides of
%transition metals and dimerized polyacetylene chains.

\medskip

\noindent DOI: 10.1103/PhysRevA.74.052308 \hfill {\bf PACS}
number(s): 03.67.Hk, 03.65.Ud, 71.10.Fd
 %Quantum communication, Entanglement and quantum nonlocality, Lattice fermion models
\end{abstract}
%,,,,,,,,,,,,,,,,,,,,,,,,,,,,,,,,,,,,,,,,,,,,,,,,,,,,,,,,,,,,,,,,,,,,,,,,,,,,,,

\vspace*{10pt}
%\keywords{The contents of the keywords}

%{\bf PACS} numbers: 05.45.-a, 87.10.+e}  % PACS, the Physics and Astronomy
%DOI: 10.1103/PhysRevA.74.052308

\maketitle

    %\section{Enhancement of Entanglement}

%\minitoc \clearpage

\section{Introduction}
Entanglement has been described as an important characteristic for
quantum information and quantum computation \cite{nc2000}.  In the
following, the quantity of interest will be the concurrence measure
of entanglement introduced by Wootters \cite{wootters1}. It will be
studied for various kinds of condensed-matter lattice systems and
the main focus will be on calculating optimized entangled states
using genetic algorithms \cite{genschemas1}. To the best of our
knowledge, at the present time there exists only a paper by Prashant
on the application of genetic algorithms to evolving quantum
circuits \cite{prashant}.

The motivation for our work resides in recent studies about maximum
nearest-neighbor entanglement \cite{romls31, romls32}. In these
cases, a $N$-qubit ring in a translationally invariant quantum state
has been considered. Under certain conditions, O'Connor and Wooters
\cite{romls31} have found formulas to obtain the maximum possible
nearest-neighbor entanglement. Moreover, they have compared this
quantity with the entanglement produced off an antiferromagnetic
state of a ring with an even number of spin-$\frac{1}{2}$ particles.
Also, there have been studies of concurrence for nearest-neighbors
in finite clusters with the purpose to see its behavior in two
dimensions. In particular, this was carried on for square,
triangular and Kagom\'e lattices \cite{romls33}.  Further studies
focus on systems with higher order entanglement, that is, when
subsystem $A$ is bigger than two qubits \cite{br01}.

In the lattice systems of this research, qubits are represented by
sites in a lattice or a chain. The two computational basis $|0\rangle$
and $|1\rangle$ are represented by occupied and empty sites,
respectively.  Using this representation, the concurrence can be
calculated for different fillings. This approach might be useful for
physical experiments involving electron control (e.g., quantum dots
\cite{recipes-spinbased}).

The electronic system will be described by a tight-binding Hamiltonian
of the form
\begin{equation}
\label{eq:hamiltonianoTB}
    \hat{H} = \sum_i \varepsilon_i \hat{n}_i + \sum_{\langle ij \rangle} t_{ij} \hat{c}_i^\dagger
    \hat{c}_j~,
\end{equation}
where, for simplicity, we will consider spinless electrons. In
\eqref{eq:hamiltonianoTB} $\hat{c}_i^\dagger$ ($\hat{c}_j$) is the
usual creation (annihilation) operator of a spinless electron at site
$i$, whereas $\hat{n}_i = \hat{c}_i^\dagger \hat{c}_i$ is the number
operator, and $t_{ij}$ is the hopping integral between
nearest-neighbors (NN) and next-nearest-neighbors (NNN) sites $i$ and
$j$. $\varepsilon_i$ is the on-site energy for atom $i$. For
simplicity, we work with the same kind of atoms and we take
$\varepsilon_i = 0$.  There exist physical systems which can be
modeled using a similar approach as here, most notably the
polyacetylene systems \cite{Su80} in the static approximation (without
phonons).

%{\it
Notice that we are using a one-body Hamiltonian.  For this case
the ground state wave function can be obtained simply by filling the
lowest one-body eigenstates. The corresponding eigenvalues are
obtained by diagonalizing the matrix Hamiltonian as expressed in the
one-body basis. In this case, one can always find a linear
transformation of $c_i^{\dagger}$ ($c_i$) leading to a diagonal form
of the Hamiltonian.

Specifically, our tight-binding Hamiltonian $\hat H = -\sum_{ij}
t_{ij} \hat c_{i}^{\dagger} \hat c_j$ can be written as $\hat H=
\sum_k \varepsilon_k \hat c_{k}^{\dagger} \hat c_k$, where
$c_i^{\dagger}= \sum_k \alpha_{ik} \hat c_{k}^{\dagger}$ and $c_i =
\sum_k \alpha_{ik}^* \hat c_{k}$. For periodic systems with only NN
hopping integrals ($t_{ij}=t$) \cite{Zan}, the Hamiltonian $\hat H$
can be diagonalized through the Fourier transformations $\hat c_l=
\frac{1}{\sqrt N} \sum_{k=1} ^{N} \exp (i 2 \pi lk/ N) \hat c_k$ and
the complex-conjugated counterpart, therefore $\hat H$ takes the
form
\begin{equation}\label{hper}
\hat H =-2t \sum_{k=1} ^{N} \cos (2 \pi k/N)
\hat c_{k}^{\dagger} \hat c_{k}~.
\end{equation}
On the other hand, for a nonperiodic  system with arbitrary hopping
integrals it is not possible to get an analytic diagonalization
procedure. For the simplest nonanalytic case of a system of $N$
atomic sites and nearest-neighbor hopping integrals, one should
diagonalize a Hamiltonian matrix of tridiagonal form:
%de hopping con valores arbitrarios no es posible una
%solucion analitica y es necesario resolver el problema de manera
%numerica. Por ejemplo, para el caso de un sistema con $N$ sitios
%atomicos y con  integrales de hopping  a primeros vecinos es
%necesario diagonalizar la matriz hamiltoniana dada por
\begin{equation}\label{hnonper}
   \hat  H = \left(
    \begin{array}{ccccccc}
        0 & -t_{12} & 0 & 0 & \cdots & \cdots &0 \\
        -t_{12} & 0 & -t_{23} &  0 & \cdots & \cdots & 0 \\
        0 & -t_{23} & 0 & -t_{34} & \cdots & \cdots & 0  \\
        0 & 0  & -t_{34} & 0 &  -t_{45} &  \cdots & 0  \\
    \cdots & \cdots & \cdots & \cdots & \cdots & \cdots & \cdots \\
    0 & 0 & \cdots & \cdots & \cdots & 0 &  -t_{(N-1)N}\\
    0 & 0 & \cdots & \cdots & 0&   -t_{(N-1)N} & 0
    \end{array} \right)~.
\end{equation}
We have used the LAPack subroutines \cite{lapack} to diagonalize
this type of Hamiltonian matrices through the QR algorithm as well
as more complicated forms resulting from next-nearest interactions,
where the Householder reduction to tridiagonal forms is applied
first.
%by using the LAPack subroutines \cite{lapack}.
The eigenvectors for $K$ fermions can be obtained in a direct way
using the preceding equations and are given by the tensorial product
of the one-body eigenvectors,
\begin{equation}\label{eigenv}
|{\bf k}_K \rangle =   \hat c_{k _1}^{\dagger}
\hat c_{k _2}^{\dagger} \cdots  \hat c_{k _K}^{\dagger}
|0 \rangle ~.
\end{equation}
Thus, it is clear that for the ground state we require the lowest
levels to be occupied.
%}

Moreover, through a Jordan-Wigner transformation the spinless
fermion Hamiltonian can be rewritten as a $XX$ spin-$\frac{1}{2}$
chain. In the spinless fermion case, each lattice site is either
occupied or free, whereas in the spin polarization case each lattice
site can have the spin up or down.
%In this way, all our work can be related in direct
%way with a XX spin 1/2 chain.
It is well known that oxides and fluorides of transition metals,
e.g., MnO, NiO and MnF$_2$, FeF$_2$, CoF$_2$, respectively, are
described by such simple spin Hamiltonians \cite{gpp03}.

Wootters' formula %\eqref{eq:wottersformula2}
 that we used to calculate the concurrence is
obtained in the Appendix and is given by
    %First, we have to obtain the density matrix $\rho_A$ for qubits $i,j$.

\begin{equation}\label{final_conc}
C = \max \{ 0, 2|\rho_{23}| - 2\sqrt{\rho_{11}\rho_{44}}\}=2\, \max
\{ 0, |\rho_{23}| - \sqrt{\rho_{11}\rho_{44}}\}~.
\end{equation}
The concurrence calculations in Sec. II are always a sum over all
the pairs of sites divided to the total number of sites. Section III
contains some conclusions and the Appendix is devoted to the
mathematics of the concurrence formula.

\section{Optimizing Entanglement using Genetic Algorithms}   %Section 2

    %There have been recent studies about maximum nearest-neighbor entanglement \cite{romls31, romls32}. In these cases, a $N$ qubit ring in a
    %translationally invariant quantum state has been considered. Under certain conditions, O'Connor and Wooters \cite{romls31} have found formulas
    %to obtain the
    %maximum possible nearest-neighbor entanglement. Moreover, they have compared this quantity with the entanglement produced off an antiferromagnetic
    %state of a ring with an even number of spin $1/2$ particles.

    %Also, there have been studies of concurrence for nearest-neighbors in finite clusters with the purpose to see its behavior in two dimensions.
    %In particular,
    %this was carried on for square, triangular and Kagom\'e lattices \cite{romls33}.

    %Further studies focus on systems with higher order entanglement, that is, when subsystem $A$ is bigger than two qubits \cite{br01}.

We pass now to the main goal of the paper which is the maximization of
entanglement using genetic algorithms. Specifically, we will consider
the ground state of a spinless system modeled by the tight binding
Hamiltonian given in Eq.~\eqref{eq:hamiltonianoTB}.

We recall that genetic algorithms (GAs) were invented by John Henry
Holland in the 1960s and were developed by him and his students and
colleagues at the University of Michigan in the 1960s and the
1970s. Holland's goal was not to design algorithms to solve specific
problems, but rather to formally study the phenomenon of adaptation as
it occurs in nature and to develop ways in which the mechanisms of
natural adaptation might be imported into computer systems.

Much alike nature, Holland's GA is a method for moving from one
population of ``chromosomes'' (e.g., strings of characters or
numbers) to a new population by using a kind of ``natural
selection'' together with the genetics-inspired operators of
crossover, mutation, and inversion (this last operator is rarely
used nowadays).

    The genetic pseudoalgorithm employed by us here goes as follows:
\begin{enumerate}
    \item Read input parameters including type of lattice, sites in the system, number of generations, crossover probability, mutation
    probability, etc.
    \item Build a table with indices of the nearest neighbors of each site. A table including also next-nearest neighbors can be built as well.
    \item Using the neighbor table, identify the specific places in the Hamiltonian matrix where ``bonds'' occur. Each place
    represents
    a valid $t_{ij}$ entries and will be stored in a special array. This array will be considered hereafter as a {\em chromosome}.
    \item Allocate two arrays, ``generation0'' and ``generation1'' composed of chromosomes.
    \item Construct an additional chromosome called ``best'' with initial random numbers between $(0,5)$.
    \item For a given range of filling repeat:
    \begin{itemize}
        \item Initialize ``generation0'' with random values in the range $( 0,5 )$.
        \item Make the first chromosome of ``generation0'' equal to ``best''.
        \item For a given number of generations repeat:
        \begin{itemize}
            \item Decode each chromosome in ``generation0'' into a Hamiltonian matrix, diagonalize it and calculate the total concurrence between
            all nearest neighbors of the system. In other words, calculate fitness for each individual in ``generation0''.
            \item Make ``best'' equal to the chromosome with highest value of fitness in ``generation0''
            \item Print the value of the average fitness of the population of ``generation0'' and fitness of ``best'' in output files.
            \item Apply the selection operator: Use crossover and mutation operators on chromosomes in ``generation0'' to create new chromosomes
            into ``generation1''.
            \item Make ``generation0'' equal to ``generation1''.
            \item Make the first chromosome in ``generation0'' equal to ``best''.
        \end{itemize}
        \item Find the chromosome with the maximum fitness. Print its fitness value in an output file.
    \end{itemize}
\end{enumerate}

To make the calculations tractable the biggest system that we
considered was of 49 sites with 800 generations for which the
optimization procedure has taken about three days. In all
calculations we have worked with the probability of crossover $p_c =
0.70$ and the probability of mutation $p_m =  0.002$.

\subsection{One-dimensional chains}

\begin{figure}[htb]
\begin{center}
\resizebox{125mm}{!}{\includegraphics{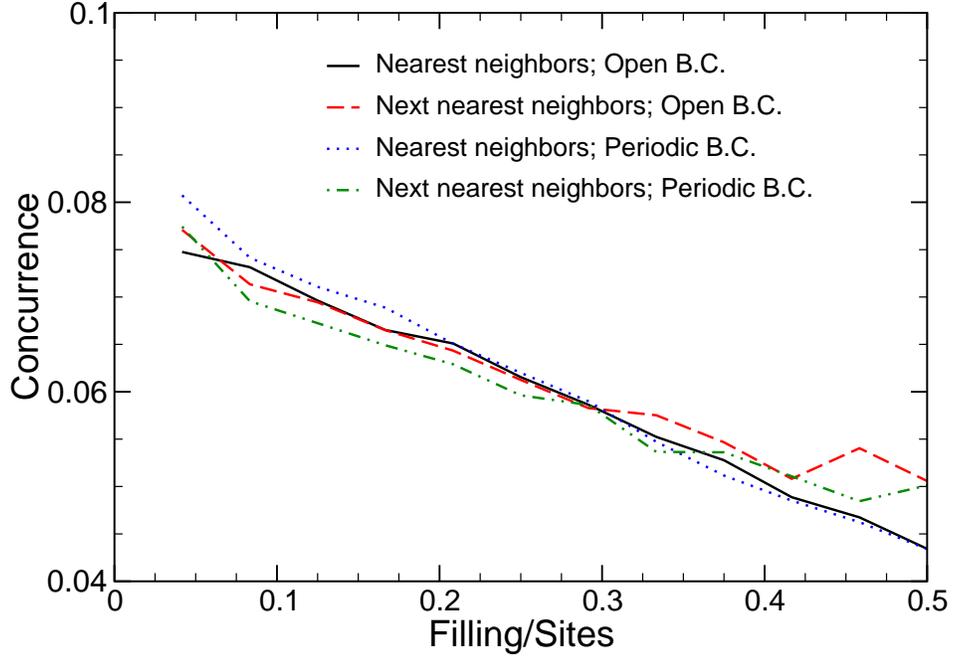}}
\caption{(Color online) Optimized concurrence as a function of band
filling for a linear chain with 24 sites. $250$ generations;
population $=400$. Note that in all figures the quantities plotted
are dimensionless and the values of the parameters $p_c$ and $p_m$
are fixed at $0.70$ and $0.002$, respectively.}
\label{fig:1-024-x-x-250-400}
\end{center}
\end{figure}

\begin{figure}[htb]
\begin{center}
\resizebox{125mm}{!}{\includegraphics{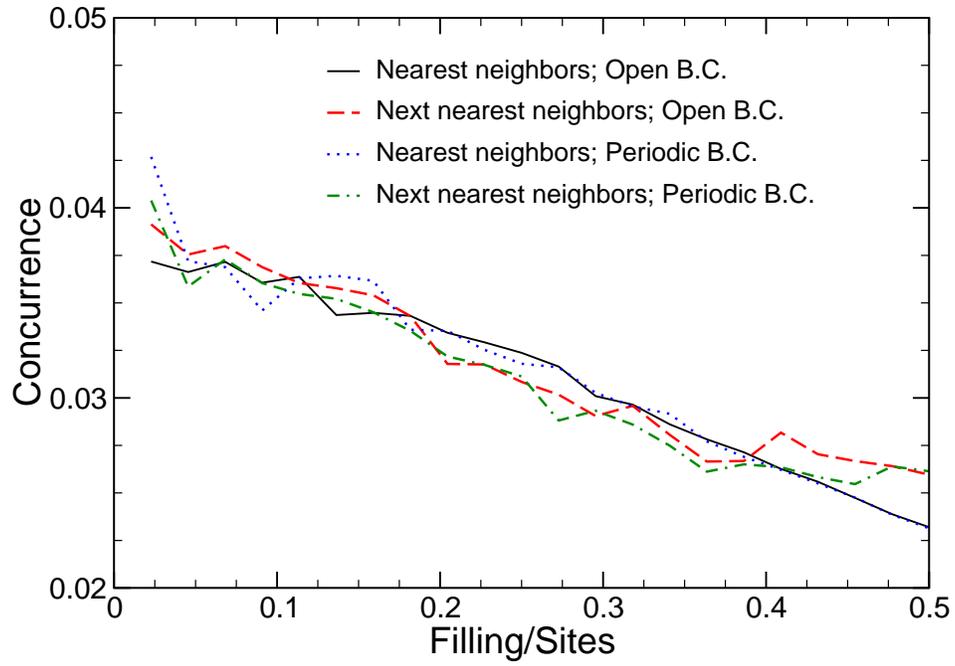}}
\caption{(Color online) Optimized concurrence as a function of band
filling for a linear chain with 44 sites. $250$ generations;
population $=400$.} %$p_c=0.70$; $p_m=0.002$.}
\label{fig:1-044-x-x-250-400}
\end{center}
\end{figure}

    We begin the analysis of entanglement maximization using genetic algorithms with the simplest case of small lineal chains with and without periodic
    boundary conditions. In Figs. \ref{fig:1-024-x-x-250-400} and \ref{fig:1-044-x-x-250-400} we present results of concurrence as a function of
    percentage filling for two chains with $24$ and $44$ sites, respectively. Besides nearest-neighbor interactions, we have also considered
interactions with both nearest neighbors and next-nearest
    neighbors in the Hamiltonian. The population size remained at $400$ individuals and the generations were kept at $250$. Later on, the role of
    the number of generations will become apparent.

    From the figures, it can be noticed that in the case of nearest-neighbor interactions with and without periodic boundary conditions,
    concurrence as function of filling is smoother than in the cases where next nearest neighbors are also considered. This can be due to a larger
    size in the chromosomes in the latter case and a greater number of generations are necessary to obtain a similar behavior than its
    only-nearest-neighbors counterpart. We can only conclude that a greater number of
generations and possibly a greater size in the population is
necessary to
    overcome these oscillations.

    Also, notice that cases including next nearest neighbors cannot yield lower results than the only nearest neighbors case. This is because
    the chromosomes from the former case contain the chromosomes of the latter (i.e., the NN case is a subset of the NNN case), which offers the
    possibility
    to explore a wider spectrum of solutions. In the case where this extra space yielded only lower results, the best chromosomes would be those of
    the NN space. This phenomenon can be most clearly noticed near half filling. Once
again, this behavior is a consequence of a greater number of
generations.

    At this point, it is important to remember that there are various parameters responsible for a larger chromosome in this kind of system. These
    parameters are the size of the system, the periodic boundary conditions and bringing next-nearest-neighbors interactions into play. A larger
    chromosome would allow an exploration of a wider solution space but on the other hand it is expected to decrease the convergence time.

\begin{figure}[htb]
\begin{center}
\resizebox{125mm}{!}{\includegraphics{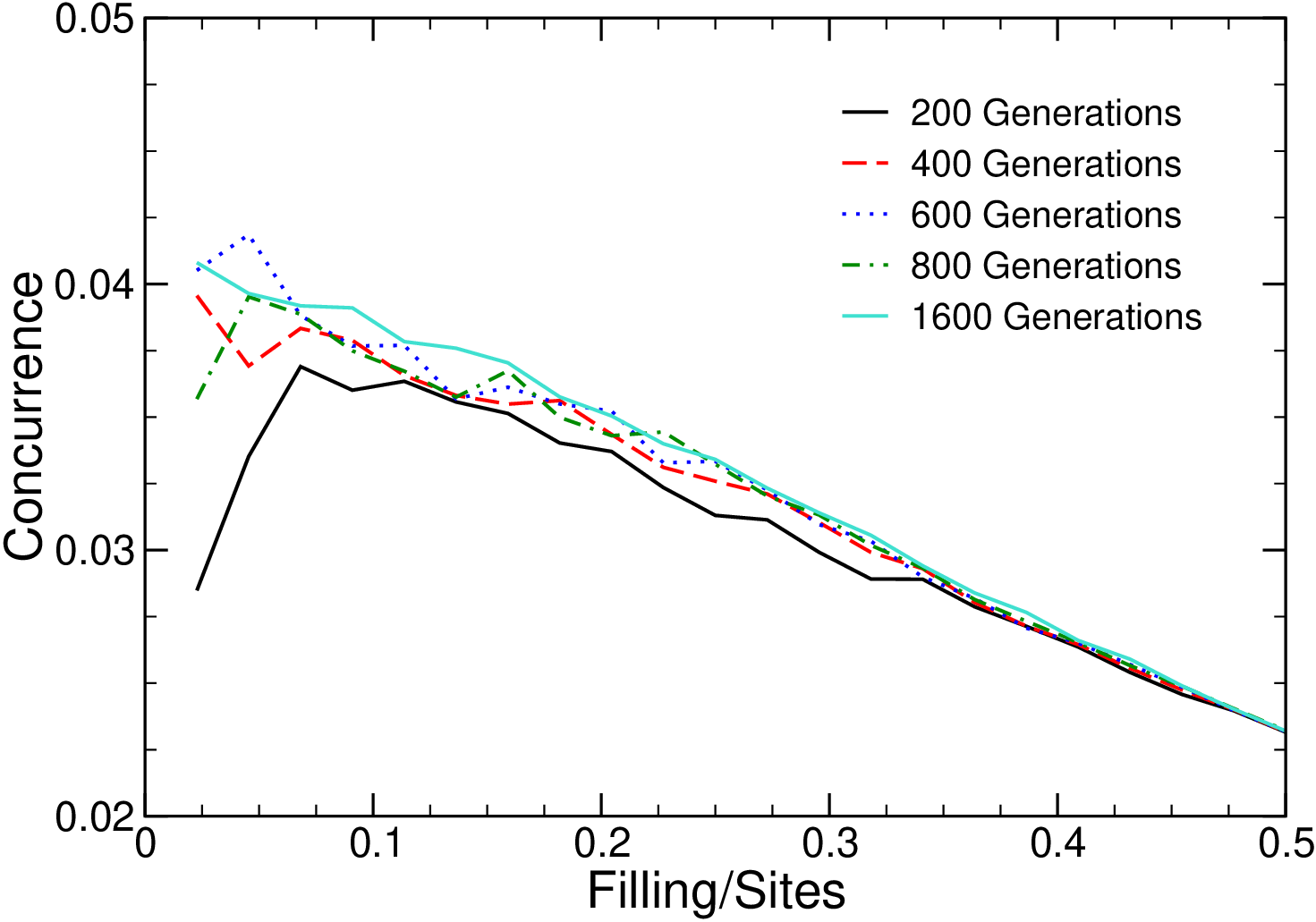}}
\caption{(Color online) Optimized concurrence for different number
of generations in a one-dimensional 44-site lattice. Open boundary
conditions, population size $= 400$.} %$p_c=0.70$; $p_m=0.002$.}
\label{fig:1-044-0-1-xxxx-400}
\end{center}
\end{figure}

\begin{figure}[htb]
\begin{center}
\resizebox{125mm}{!}{\includegraphics{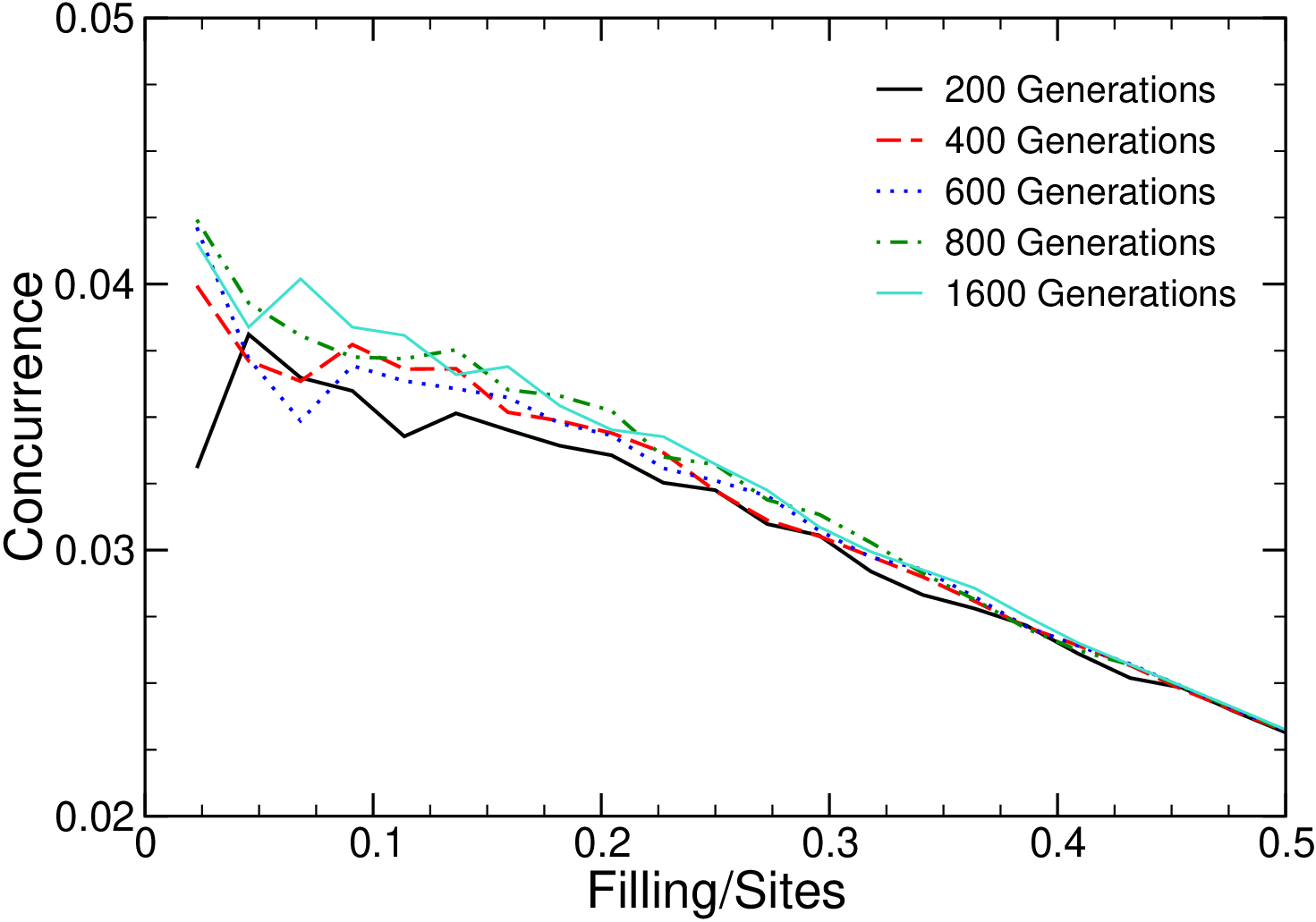}}
\caption{(Color online) Optimized concurrence for different number
of generations in a one-dimensional 44-site lattice using periodic
boundary conditions, population size $= 400$.} %$p_c=0.70$; $p_m=0.002$.}
\label{fig:1-044-1-1-xxxx-400}
\end{center}
\end{figure}

    We have already mentioned the possibility of a greater number of generations affecting directly the smoothness of the concurrence. We
    addressed this question by running two cases depicted in Figs.~\ref{fig:1-044-0-1-xxxx-400} and \ref{fig:1-044-1-1-xxxx-400}, where the former
    does not consider periodic boundary conditions while the latter does.

    In both cases we have set a $44$-site chain with a population size of $400$. Only the interactions between nearest neighbors were taken into
    account.

    Both figures confirm our early supposition about increasing the number of generations since the correlation functions look increasingly smoother.
    Notice, however, that certain roughness still remains. Some possible solutions consist of increasing the size of the population, dynamically
    change the mutation probability (when variation between individuals begins to narrow) and raising the number of generations further more.
    Different selection methods could also be considered because an inefficient parent selection could lead to slow evolution of the system. Even
    though it is clear that individuals with better fitness are obtained, notice how in Fig.~\ref{fig:1-044-0-1-xxxx-400} the best chromosome
    near $0.05$ filling was obtained with $600$ generations despite having cases with up to three times more generations. %This gives further
    %evidence about the necessity to investigate the methods discussed above.

    In Fig.~\ref{fig:Final1-044-1-1-500-350-completo}
    (for its more detailed structure up to
    $2000$ generations see Fig.~\ref{fig:Detalle1-044-1-1-500-350-completo}) we follow the evolution (optimization) of concurrence for each
    filling in a $44$-site chain with
    periodic boundary conditions and interactions only between nearest neighbors. The population size for this calculation was $350$ and the
    number of generations was $500$ per filling. %Black dots represent
    The average fitness per population is compared with the fitness from the best chromosome in the population.
    Notice how the population always follows closely the evolution of the best chromosome. Transitions between different fillings are readily
    noticed through a drop in average fitness. A very remarkable feature is that the best chromosome for a certain filling ranks high for the next
    filling but is \textsl{not} the highest. In other words, there are different best chromosomes for different band fillings.
    %Further studies are needed to determine exactly the degree of differences and their exact nature.
    We also notice that the bottom dots correspond to the average
    concurrence for randomly disordered populations and that the
    average concurrences for the subsequent
    optimized GA populations are always better than the disordered
    cases.

    Another remarkable characteristic about Fig.~\ref{fig:Final1-044-1-1-500-350-completo} is its symmetry around half-band filling.
    This property is
    due to the fact that this is a bipartite lattice and consequently its physical properties are symmetric because of an electron-hole
    transformation.
 %Remember that a lattice is considered bipartite if it can be divided into two independent lattices.
 The fact that the results presented show this property reassures the validity of our calculations.

\subsection{Two-dimensional systems}
    %Recently, the effect of the dimensionality on concurrence has begun to be studied in square, triangular and Kagom\'e lattices \cite{romls33}.
    We will study now the optimization of concurrence in two-dimensional systems modeled by means of the same tight-binding Hamiltonian.

\subsubsection{Square lattices}
\begin{figure}[htb]
\begin{center}
\resizebox{125mm}{!}{\includegraphics{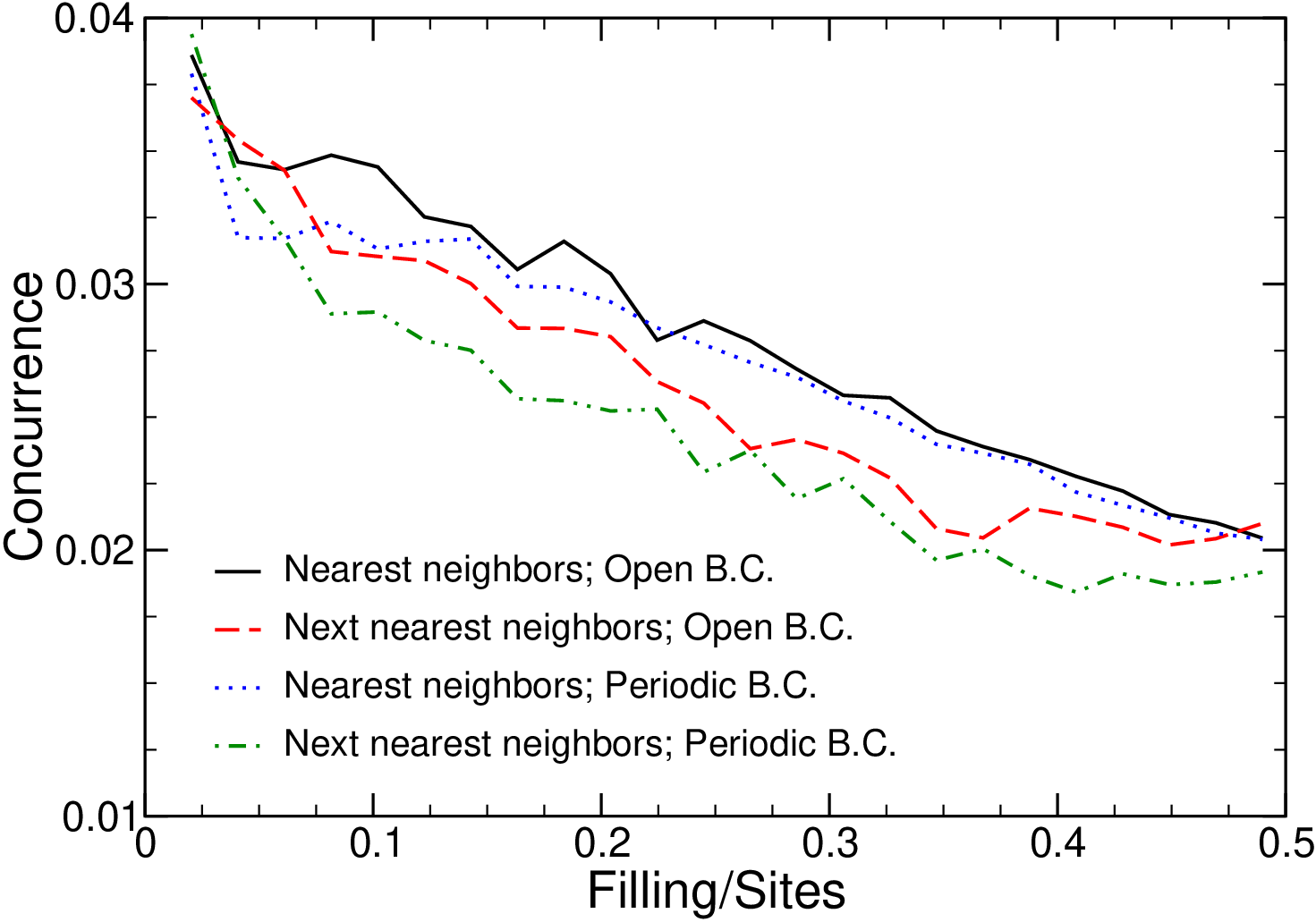}}
\caption{(Color online) Optimized concurrence for a square lattice
comparing nearest-neighbor interactions, next-nearest-neighbor
interactions and boundary conditions.
        $600$ generations; population size $= 350$.}  %$p_c=0.70$; $p_m=0.002$.}
\label{fig:2-1-07-07-x-x-600-350}
\end{center}
\end{figure}

\begin{figure}[htb]
\begin{center}
\resizebox{125mm}{!}{\includegraphics{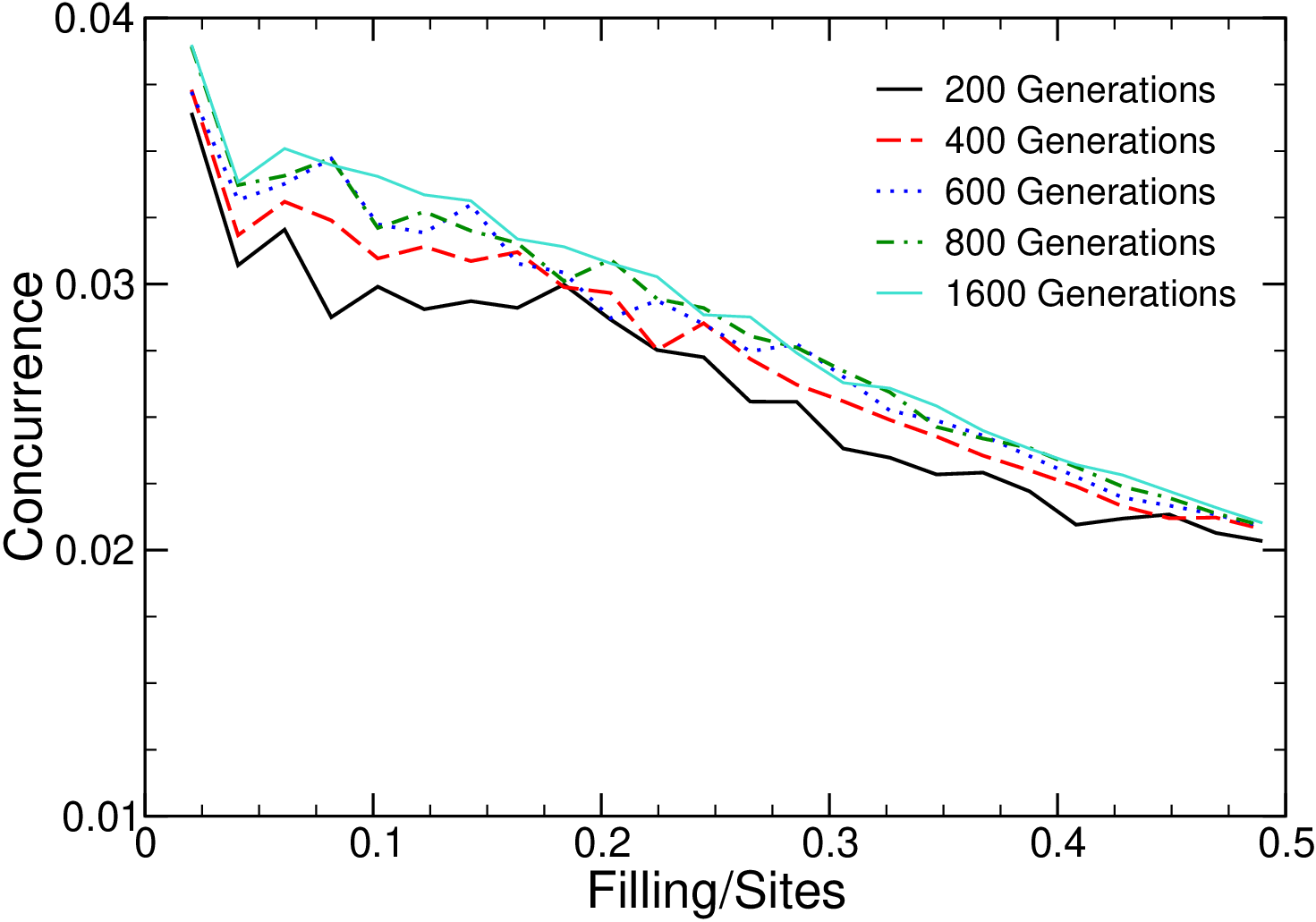}}
\caption{(Color online) Optimized concurrence for a square lattice
comparing number of generations for nearest-neighbor interactions,
periodic boundary conditions, population
        size $= 400$.} %$p_c=0.70$; $p_m=0.002$.}
\label{fig:2-1-07-07-1-1-xxxx-400}
\end{center}
\end{figure}

    In Figs.~\ref{fig:2-1-07-07-x-x-600-350} and \ref{fig:2-1-07-07-1-1-xxxx-400} we display the concurrence as a function of band filling for
    a 7$\times$7 square
    lattice.

    Figure \ref{fig:2-1-07-07-x-x-600-350} presents a comparison between systems using nearest-neighbor and next-nearest-neighbor interactions
     as well as periodic and open boundary conditions. The number of generations for these cases has been chosen $600$ and the population size $350$.
     It is worth mentioning that
     in general the cases with interactions only between nearest-neighbors rank slightly higher in its concurrence value. This is a somewhat
     unexpected result that may be attributed to various factors including selection methods and number of generations. Possible reasons for this
behavior were addressed in the preceding section.

    To study the effect of the number of generations on the optimized value of concurrence and the smoothness of the curve, we present calculations
    for four different cases in Fig.~\ref{fig:2-1-07-07-1-1-xxxx-400}. In these cases, population size was kept at $400$. It is clear that
    by raising this number we are able to obtain better optimized solutions and the concurrence curve tends to be smoother.

    %We believe it is necessary to make further investigations on the effect of the population size, as well as different selection methods.

\subsubsection{Triangular lattices}
    Finally, we have made calculations for not bipartite lattices in order to study the effect of frustration on concurrence.
    It has already been mentioned that these kinds of lattices are not symmetric under an electron-hole transformation.
    This is the reason why their physical properties differ completely between lower and upper sections of band filling.

\begin{figure}[htb]
\begin{center}
\resizebox{125mm}{!}{\includegraphics{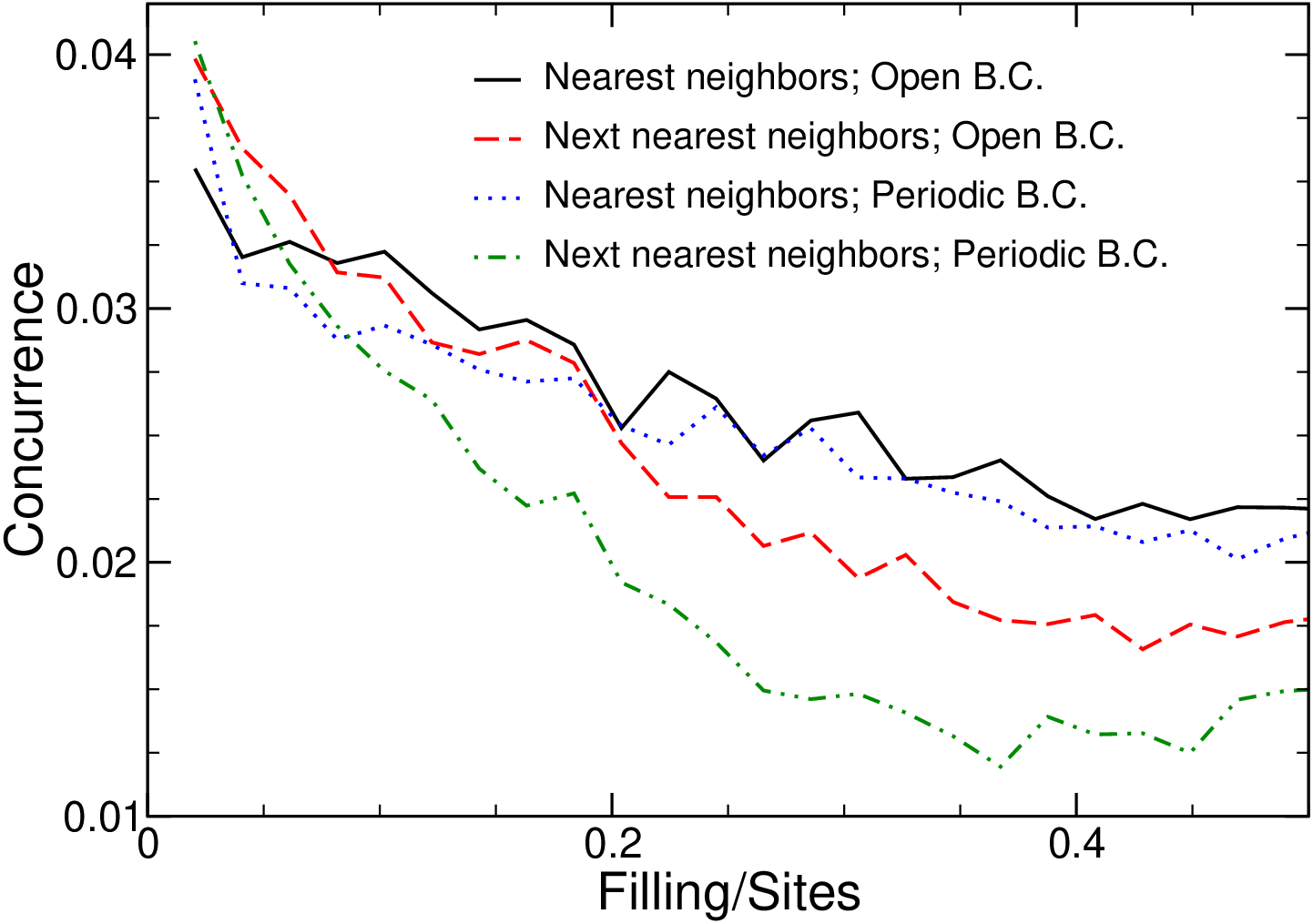}}
\caption{(Color online) Optimized concurrence for a triangular
lattice comparing nearest-neighbor interactions, next nearest
neighbor interactions and boundary conditions. $600$ generations;
population size $= 350$.} %$p_c=0.70$; $p_m=0.002$.}
\label{fig:2-2-07-07-x-x-600-350}
\end{center}
\end{figure}

\begin{figure}[htb]
\begin{center}
\resizebox{125mm}{!}{\includegraphics{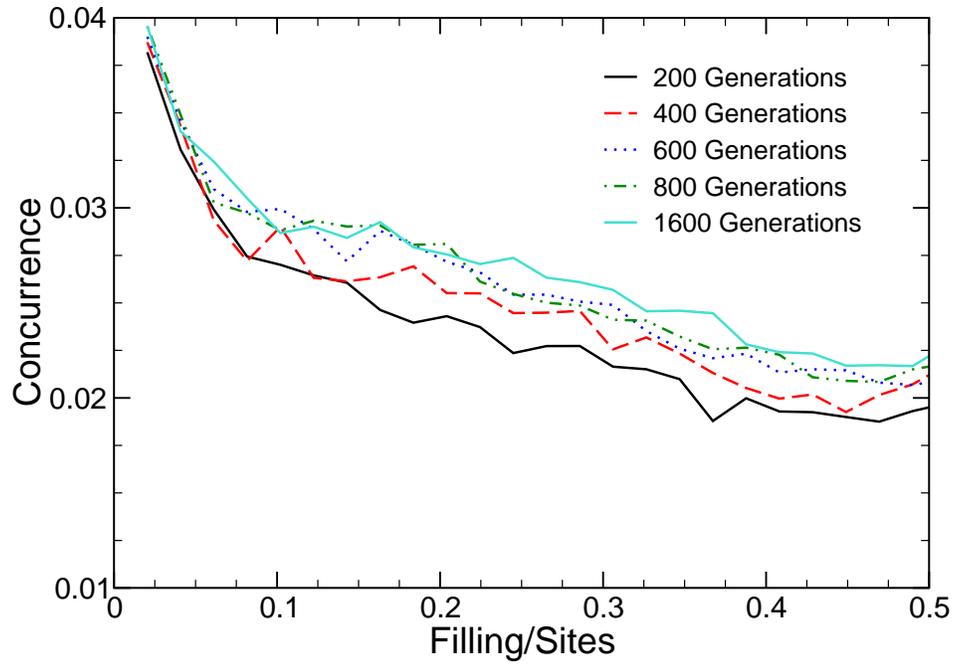}}
\caption{(Color online) Optimized concurrence for a triangular
lattice comparing number of generations for nearest-neighbor
interactions, periodic boundary conditions, population
        size $= 400$.} %$p_c=0.70$; $p_m=0.002$.}
        \label{fig:2-2-07-07-1-1-xxxx-400}
\end{center}
\end{figure}

    As a particular case of a nonbipartite lattice, we have considered a triangular lattice with $49$ sites. Results of our calculations are shown in
    Figs.~\ref{fig:2-2-07-07-x-x-600-350} and \ref{fig:2-2-07-07-1-1-xxxx-400}.

    In Fig.~\ref{fig:2-2-07-07-x-x-600-350} a population size of $350$ has been used and the system has been allowed to go up to $600$ generations.
    As in the preceding sections, this case includes the interaction between nearest and next-nearest neighbors, as well as open and periodic boundary
    conditions. Once again, we find better optimizations for nearest-neighbor interactions. It is important to remember that we are dealing with a
    more complex chromosome, as sites in this kind of lattice have a greater number of neighbors than one dimensional systems. This is also a cause
    for a lower time in convergence as the solution space increases considerably.

    The effect of the number of generations can be examined in Fig.~\ref{fig:2-2-07-07-1-1-xxxx-400}. These results demonstrate the slow convergence
    when calculating this kind of system. Notice that, although oscillations decrease and better individuals are found, efficiency narrows between
    the cases with $600$ and $800$ generations.

\section{Conclusions}
    %In this work, we present the necessary formulation to measure entanglement through concurrence for a pair of sites.
    %By doing this, we have been able to calculate the behaviour of concurrence and the effect of off-diagonal disorder in rings.
    We have implemented computational techniques --more specifically genetic algorithms-- to optimize entanglement in systems modeled according
    to a tight-binding Hamiltonian. The qubits in all these studies have been described as sites in the system and the computational basis as
     occupied or empty sites.

\begin{figure}[htb]
\begin{center}
\resizebox{125mm}{!}{\includegraphics{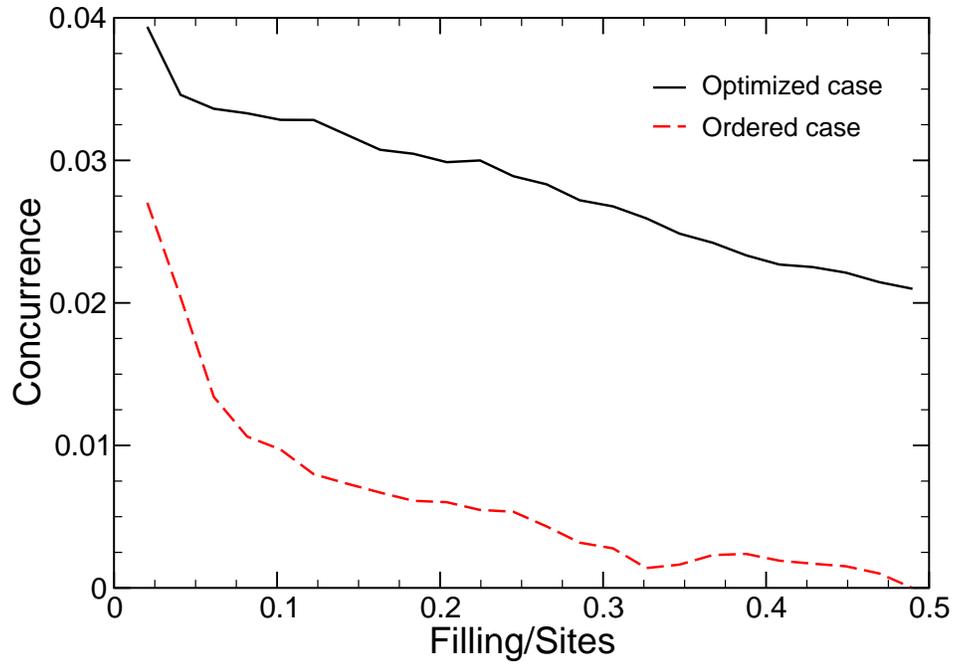}}
\caption{(Color online) Concurrence for the optimized and ordered
cases in a $49$ sites square lattice using nearest-neighbor
interactions and periodic boundary conditions $800$ generations;
population size $=400$.} %$p_c=0.70$; $p_m=0.002$.}
%$800$ generations were run for each filling and population size was $400$. Only nearest-neighbor interactions were allowed and periodic boundary conditions were used. The crossover and mutation probabilities were $p_c=0.70$ and $p_m=0.002$, respectively. The ordered case was calculated with off diagonal values of $t_{NN}=1$}
\label{fig:ordenVSoptimizedCuadrada}
\end{center}
\end{figure}

\begin{figure}[htb]
\begin{center}
\resizebox{125mm}{!}{\includegraphics{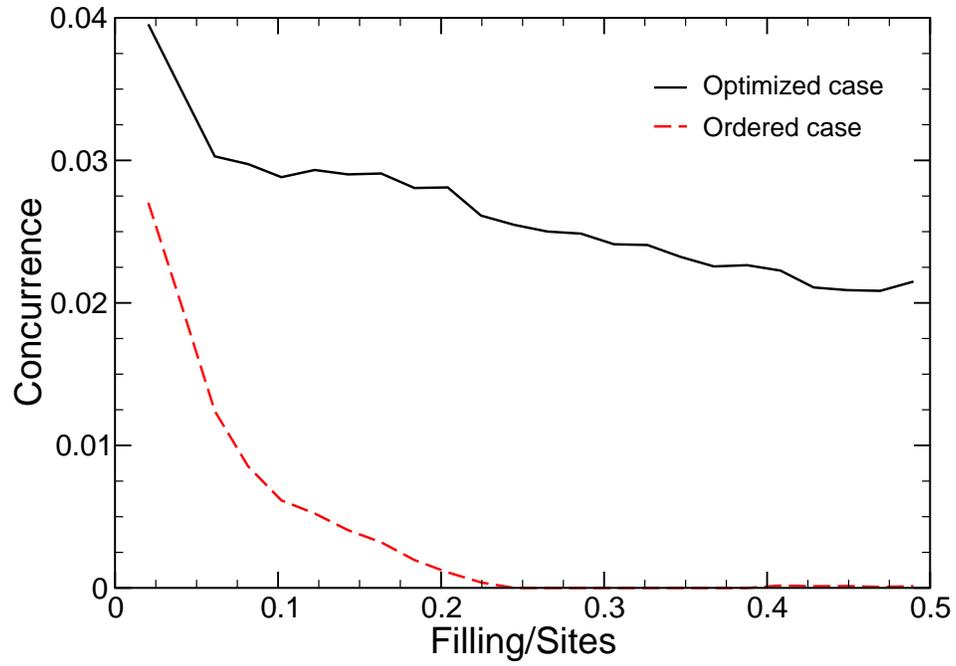}}
\caption{(Color online) Concurrence for the optimized and ordered
cases in a $49$ sites triangular lattice under the same conditions
as in the previous figure.} \label{fig:ordenVSoptimizedTriangular}
\end{center}
\end{figure}

\begin{figure}[htb]
\begin{center}
\resizebox{125mm}{!}{\includegraphics{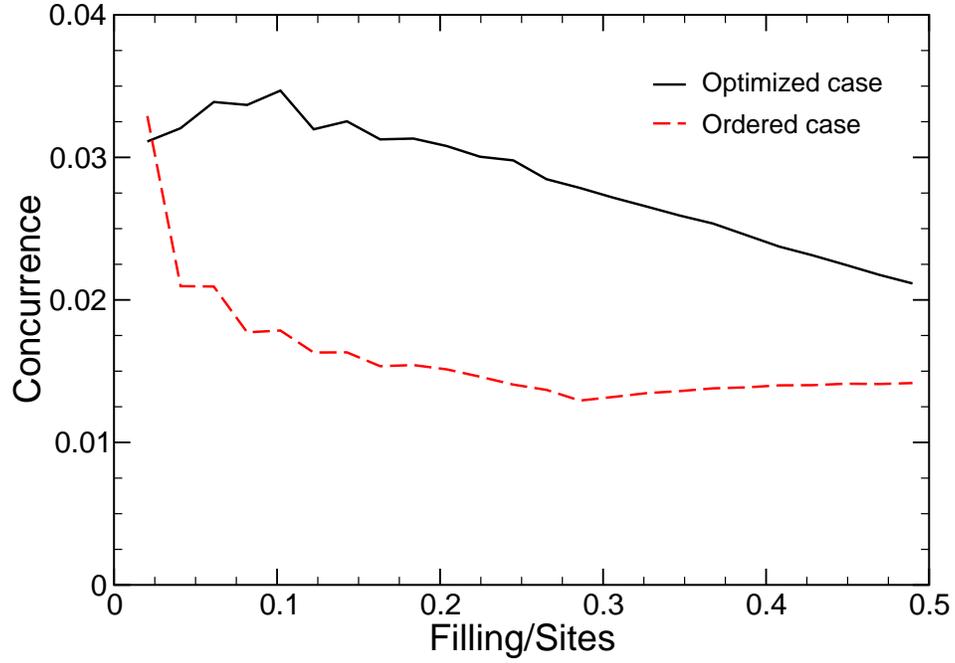}}
\caption{(Color online) Comparison between the optimized and the
ordered case in a one-dimensional system of $49$ sites. The
calculation
    parameters were identical to those of Figs.~\ref{fig:ordenVSoptimizedCuadrada} and \ref{fig:ordenVSoptimizedTriangular}.}
\label{fig:ordenVSoptimized1D}
\end{center}
\end{figure}

    Our application of genetic algorithms has proven to be valuable, since we obtained configurations which yield better results
    for concurrence in the randomly disordered one-dimensional case as discussed in Sec. II A (in fact, we have partial results confirming this
    statement for two-dimensional cases as well). Moreover, the GA
    optimization provided better results even with respect to the
    ordered cases as can be noticed in
    Figs.~\ref{fig:ordenVSoptimizedCuadrada}, \ref{fig:ordenVSoptimizedTriangular}, and \ref{fig:ordenVSoptimized1D}.

    We finally mention that the optimal Hamiltonian for the one-dimensional case
with periodic boundary conditions corresponds to dimerized chain for
which the coupling coefficients take alternate high-low values in
the half filling region. This structural phase transition has been
found in the polyacetylene chains and suggests that optimal
entanglement can be obtained in the dimerized phase of the
polyacetylene. We have also some evidence of this behavior in
two-dimensional systems. These results will be reported elsewhere.

    Quantum computation and quantum information are still a long way to go. Nevertheless, these areas seems to be a logical and necessary step in
    tomorrow's technological world. In this scenario, quantum entanglement will play a critical role, and our work attempts to be another step
    towards better understanding it.

\bigskip

\newpage

{\bf APPENDIX: THE CONCURRENCE}

\medskip

%.......................

\begin{figure}[b]
    \centerline{\includegraphics[width=3.5in]{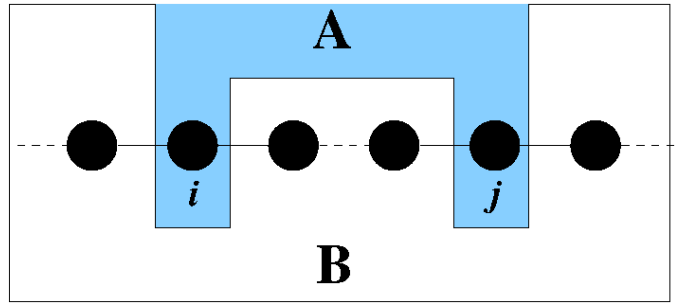}}%[scale=1.7]{scheme_conc.eps}}
    \caption{(Color online) Schematic illustration of the partition of the system of interest into two subsystems for the calculation of
    the concurrence between sites $i$ and $j$.} \label{fig:scheme}
\end{figure}

To calculate the concurrence we need first the density matrix
$\rho_A$ for qubits $i,j$, which is the trace over system $B$ of all
the possible states $|\psi_{AB}\rangle \langle \psi_{AB}|$.
    The general state function for this system is
\begin{equation}
\label{eq:generalAB}
    |\psi_{AB}\rangle = \sum_n \alpha_n |\psi_{A}\rangle|\psi_{B}\rangle
\end{equation}
    where, for a system of $N$ sites, $n$ goes through all the $2^N$ possible combinations in the computational basis
    (e.g., $|00\ldots00\rangle \rightarrow |11\ldots11\rangle$). Subsystem $A$ is comprised of the two qubits of interest in the sites $i,j$
    (i.e., $|\psi_i\rangle \otimes|\psi_j\rangle$, see also Fig.~\ref{fig:scheme}). For a specific system of $N$ sites, there are $N_1$ occupied
    and $N-N_1$ not occupied sites. Our two-qubit subsystem $A$ has, naturally, four possible states --namely $|00\rangle, |01\rangle,
    |10\rangle$,
    and $|11\rangle$-- therefore Eq.~\eqref{eq:generalAB} can be decomposed in the following
    manner:
\begin{equation}
%\begin{split}
    |\psi_{AB}\rangle = \sum_m a_m |00\rangle \otimes |\psi_B^m\rangle + \sum_o b_o |01\rangle \otimes |\psi_B^o\rangle
    + \sum_p c_p |10\rangle \otimes |\psi_B^p\rangle  + \sum_q d_q |11\rangle \otimes |\psi_B^q\rangle.
%\end{split}
\end{equation}
    In this equation, the sums run for all the possible combinations in the $|\psi_B\rangle$ space such that the number $N_1$ of occupied sites is
    preserved. For example, if $|\psi_A\rangle= |01\rangle$, system $B$ is left with $N_1 - 1$ occupied sites.
%. That is, for a certain filling $N_1$, system $A$ takes zero, one, one and two occupied sites from the whole system respectively
%(e.g. for $|\psi_A\rangle = |01\rangle$, system $B$ is left with
%$N_1 - 1$ occupied sites). In general, the sums run as
%\begin{equation}
%   no se me ocurre la ***** formula luego la pongo
%\end{equation}
%   with $L$ being system B's space ($L=N-2$) and $F$ the effective filling for system B $(F=N_1-(\hat{n}_i + \hat{n}_j)|\psi_A\rangle$).

    To obtain the reduced density matrix it is necessary to perform the trace over system
    $B$,
\begin{equation}
    \rho_A = \sum_{l=1}^{2^{N-2}} \big(\langle I |\otimes \langle \psi_B^l |\big) |\psi_{AB}\rangle \langle \psi_{AB}| \big( | I \rangle \otimes |
    \psi_B^l\rangle \big).
\end{equation}
    It is clear that applying this operation will not eliminate those terms whose elements in the $B$ subsystem in $|
    \psi_{AB} \rangle\langle \psi_{AB}|$ have the same number of occupied sites. The terms in system $A$ that are left after the trace operation
    are of the kind $|00\rangle\langle 00|$, $|01\rangle\langle 01|$, $|01\rangle\langle 10|$, $|10\rangle\langle 01|$, $|10\rangle\langle
    10|$,
    and $|11\rangle\langle 11|$.

    The $|00\rangle\langle 00|$ element is spared after the trace operator because its $|\psi_B^m\rangle$ elements contain the same quantity of
    occupied sites (i.e., $N_1$ sites). This is a similar case with the $|11\rangle\langle 11|$ elements where the $|\psi_B^q\rangle$ wave
    functions contain $N_1 - 2$ occupied sites.

    In the case of $|01\rangle\langle 01|$, $|01\rangle\langle 10|$, $|10\rangle\langle 01|$, and $|10\rangle\langle 10|$ elements, notice
    how their $|\psi_B^{o,p}\rangle$ wave functions have the same number of occupied sites ($N_1 - 1$).

    Finally, the elements in the reduced density matrix are
\begin{equation}
    \rho_A = \left(
    \begin{array}{cccc}
        \rho_{11} & 0 & 0 & 0 \\
        0 & \rho_{22} & \rho_{23} & 0 \\
        0 & \rho_{32} & \rho_{33} & 0 \\
        0 & 0 & 0 & \rho_{44}
    \end{array} \right)~.
\end{equation}

    For $\rho_A$ to be a valid density matrix, it must be Hermitic ($\rho_A = \rho_A^{\dagger *}$) and its trace be equal to $1$.
    This means that $\rho_{32}=\rho_{23}^*$ and $\rho_{11} + \rho_{22}+ \rho_{33}+ \rho_{44} = 1$ so it is necessary to calculate only four
    elements of the matrix.

%\subsection{The density matrix elements}
    In order to calculate each of these reduced density matrix elements, the second quantization approach will be used.

    The first element of the matrix, $\rho_{11}$ can be realized as follows
\begin{equation}
    \rho_{11} = \langle \psi_{AB}| \big( 1-\hat{n}_i \big) \big( 1-\hat{n}_j \big) |\psi_{AB}\rangle ~,
\end{equation}
    where the operator $\hat{n}_j$ projects on all the elements of the type $|x1\rangle \otimes |\psi_B\rangle$ and after applying $( 1-\hat{n}_j)$ we
    end up with all the elements that do \textsl{not} occupy the site $j$ (i.e., $|x0\rangle \otimes |\psi_B\rangle$). A similar approach
    follows $( 1-\hat{n}_i )$ and after applying the bra operation we are left only with the coefficients of all the $|00\rangle \otimes |
    \psi_B\rangle$ states.

    The other elements are obtained likewise with the following operators:
    $\rho_{22} =  \langle \psi_{AB}| \big( 1-\hat{n}_i \big) \hat{n}_j
    |\psi_{AB}\rangle$,
    $\rho_{33} =  \langle \psi_{AB}| \hat{n}_i \big( 1-\hat{n}_j \big)
    |\psi_{AB}\rangle$,
    $\rho_{44} =  \langle \psi_{AB}| \hat{n}_i \hat{n}_j
    |\psi_{AB}\rangle$, and
    $\rho_{23} =  \langle \psi_{AB}| c_j c_i^\dagger |\psi_{AB}\rangle$.

    In the matrix element $\rho _{23}$, $c_i^\dagger$ maintains only those states of the form $|0x\rangle \otimes |\psi_B\rangle$ transforming them
    into $|1x\rangle \otimes |\psi_B\rangle$. Out of this set of states, $c_j$ deletes all states of the type $|x0\rangle \otimes |\psi_B\rangle$
    and we end up with states $|10\rangle \otimes |\psi_B\rangle$.
%\textcolor{red}{me dan ganas de poner un ejemplo en algun apendice para clarificar porque es dificil explicar}

    It is very easy to show that the $\rho_A$ elements can be calculated as average quantities of the complete ground-state wave function.
    For example,
%\begin{equation}
%\begin{split}
%    \rho_{11} = &\langle \psi_{AB}|\psi_{AB} \rangle - \langle \psi_{AB}|\hat{n}_i|\psi_{AB} \rangle - \langle \psi_{AB}|\hat{n}_j|\psi_{AB} \rangle
%\\ &+ \langle \psi_{AB}|\hat{n}_i\hat{n}_j|\psi_{AB} \rangle \\
%    = &1 - \langle \hat{n}_i \rangle - \langle \hat{n}_j \rangle + \langle \hat{n}_i\hat{n}_j \rangle.
%\end{split}
%\end{equation}
\begin{equation}
%\begin{split}
    \rho_{11} = \langle \psi_{AB}|\psi_{AB} \rangle - \langle \psi_{AB}|\hat{n}_i|\psi_{AB} \rangle - \langle \psi_{AB}|\hat{n}_j|\psi_{AB} \rangle
    + \langle \psi_{AB}|\hat{n}_i\hat{n}_j|\psi_{AB} \rangle \\
    = 1 - \langle \hat{n}_i \rangle - \langle \hat{n}_j \rangle + \langle \hat{n}_i\hat{n}_j \rangle.
%\end{split}
\end{equation}
The other elements are obtained similarly, $ \rho_{22} = \langle
\hat{n}_j \rangle - \langle \hat{n}_i\hat{n}_j \rangle$, $\rho_{33}
= \langle \hat{n}_i \rangle - \langle \hat{n}_i\hat{n}_j
    \rangle$, $\rho_{44} = \langle \hat{n}_i\hat{n}_j \rangle$, $\rho_{23} = \langle
c_j c_i^\dagger \rangle$.
%\begin{equation}
%\begin{array}{ll}
%    \rho_{22} = \langle \hat{n}_j \rangle - \langle \hat{n}_i\hat{n}_j \rangle & \rho_{33} = \langle \hat{n}_i \rangle - \langle \hat{n}_i\hat{n}_j
%    \rangle \\
%    \rho_{44} = \langle \hat{n}_i\hat{n}_j \rangle & \rho_{23} = \langle c_j c_i^\dagger \rangle
%\end{array}
%\end{equation}

Concurrence is an \textsl{entanglement monotone} in its own right
    (i.e., positive or zero for any density matrix $\rho$; $0$ for factorizable states and $1$ for the Bell
    states). A simple formula for the concurrence has been worked
    out by Wootters in 1998 \cite{wootters1},
\begin{equation}
\label{eq:wottersformula2}
    C(\rho) = \textrm{max}\{ 0, \lambda_1 - \lambda_2 - \lambda_3 -
    \lambda_4\},
\end{equation}
where the $\lambda$ coefficients are the square roots of the
eigenvalues of the non-Hermitian matrix $\rho_A\tilde{\rho_A}$ in
    decreasing order. The formula applies for the density matrix of the subsystem with the pair of qubits [$\rho_A=$tr$_B$($\rho$)].
The density matrix $\tilde{\rho}$ is defined through a \textsl{spin
flip} transformation expressed
    in terms of the imaginary Pauli matrix $\sigma_y$ as follows
\begin{equation}
\label{eq:spinfliprho}
    \tilde{\rho_A}=\big( \sigma_y \otimes \sigma_y \big) \rho_A^* \big( \sigma_y \otimes \sigma_y
    \big).
\end{equation}
This leads to
% In order to use the concurrence formula \eqref{eq:wottersformula2}, the non-Hermitian matrix $\rho_A\tilde{\rho_A}$ must be obtained.
    %The matrix $\tilde{\rho_A}$ is constructed using Eq.~(\eqref{eq:spinfliprho}):
\begin{equation}\label{nonH}
    \tilde{\rho_a} = \left(
    \begin{array}{cccc}
        0 & 0 & 0 & -1 \\
        0 & 0 & 1 & 0 \\
        0 & 1 & 0 & 0 \\
        -1 & 0 & 0 & 0
    \end{array} \right) \left(
    \begin{array}{cccc}
        \rho_{11}^* & 0 & 0 & 0 \\
        0 & \rho_{22}^* & \rho_{23}^* & 0 \\
        0 & \rho_{32}^* & \rho_{33}^* & 0 \\
        0 & 0 & 0 & \rho_{44}^*
    \end{array} \right) \left(
    \begin{array}{cccc}
        0 & 0 & 0 & -1 \\
        0 & 0 & 1 & 0 \\
        0 & 1 & 0 & 0 \\
        -1 & 0 & 0 & 0
    \end{array} \right) \\
  %  = &\left(
   % \begin{array}{cccc}
   %     0 & 0 & 0 & -\rho_{44}^* \\
   %     0 & \rho_{32}^* & \rho_{33}^* & 0 \\
   %     0 & \rho_{22}^* & \rho_{23}^* & 0 \\
   %     -\rho_{11}^* & 0 & 0 & 0
   % \end{array} \right) \left(
   % \begin{array}{cccc}
   %     0 & 0 & 0 & -1 \\
   %     0 & 0 & 1 & 0 \\
   %     0 & 1 & 0 & 0 \\
   %     -1 & 0 & 0 & 0
   % \end{array} \right) \\
    = \left(
    \begin{array}{cccc}
        \rho_{44}^* & 0 & 0 & 0 \\
        0 & \rho_{33}^* & \rho_{32}^* & 0 \\
        0 & \rho_{23}^* & \rho_{22}^* & 0 \\
        0 & 0 & 0 & \rho_{11}^*
    \end{array} \right)~.
\end{equation}

    Now we are able to construct the non-Hermitian matrix $\rho_A \tilde{\rho_A}$
\begin{equation}\label{rhorhot}
  %  \rho_A\tilde{\rho_A} & = &\left(
  %  \begin{array}{cccc}
  %      \rho_{11} & 0 & 0 & 0 \\
  %      0 & \rho_{22} & \rho_{23} & 0 \\
  %      0 & \rho_{32} & \rho_{33} & 0 \\
  %      0 & 0 & 0 & \rho_{44}
  %  \end{array} \right) \left(
  %  \begin{array}{cccc}
  %      \rho_{44}^* & 0 & 0 & 0 \\
  %      0 & \rho_{33}^* & \rho_{32}^* & 0 \\
  %      0 & \rho_{23}^* & \rho_{22}^* & 0 \\
  %      0 & 0 & 0 & \rho_{11}^*
  %  \end{array} \right) \\
   \rho_A\tilde{\rho_A}  =  \left(
    \begin{array}{cccc}
        \rho_{11}\rho_{44}^* & 0 & 0 & 0 \\
        0 & \rho_{22}\rho_{33}^*+\rho_{23}\rho_{23}^* & \rho_{22}\rho_{32}^*+\rho_{23}\rho_{22}^* & 0 \\
        0 & \rho_{32}\rho_{33}^* + \rho_{33}\rho_{23}^* & \rho_{32}\rho_{32}^* + \rho_{33}\rho_{22}^* & 0 \\
        0 & 0 & 0 & \rho_{11}^*\rho_{44}
    \end{array} \right)~.
\end{equation}
    However, $\rho_A$ is indeed Hermitian so the following relationships are taken into account:
    $\rho_{11} = \rho_{11}^*$, $\rho_{22}=\rho_{22}^*$, $\rho_{32}=\rho_{23}^*$, $\rho_{33}=\rho_{33}^*$, and $\rho_{44}=\rho_{44}^*$.
    Therefore, the matrix $\rho_A \tilde{\rho_A}$ has the form
%................................
\begin{equation}\label{finalrhorhot}
\rho_A\tilde{\rho_A}  = \left(
    \begin{array}{cccc}
        \rho_{11}\rho_{44} & 0 & 0 & 0 \\
        0 & \rho_{22}\rho_{33}+|\rho_{23}|^2 & 2\rho_{22}\rho_{23} & 0 \\
        0 & 2\rho_{33}\rho_{23}^* & \rho_{22}\rho_{33} + |\rho_{23}|^2 & 0 \\
        0 & 0 & 0 & \rho_{11}\rho_{44}
    \end{array} \right).
\end{equation}
%....................

    In a block diagonal matrix, the eigenvalues are simply the eigenvalues of individual blocks, so two eigenvalues are readily available.
    The other two are obtained from the following determinant
\begin{equation}
    \left( \begin{array}{cc}

        \rho_{22}\rho_{33}+|\rho_{23}|^2 - \lambda & 2\rho_{22}\rho_{23} \\
        2\rho_{33}\rho_{23}^* & \rho_{22}\rho_{33} + |\rho_{23}|^2 -\lambda \\
    \end{array} \right).
\end{equation}
    %which is
%\begin{equation}
 %   (\rho_{22}\rho_{33}+|\rho_{23}|^2 - \lambda)^2 - 4\rho_{22}\rho_{33}|\rho_{23}|^2 = 0.
%\end{equation}
    This gives %And now we simply find the value of $\lambda$
%\begin{align}
%    (\rho_{22}\rho_{33}+|\rho_{23}|^2 - \lambda)^2 = 4\rho_{22}\rho_{33}|\rho_{23}|^2 \\
%    \rho_{22}\rho_{33}+|\rho_{23}|^2 - \lambda = \pm 2 \sqrt{\rho_{22}\rho_{33}}|\rho_{23}| \\
%....................
    \begin{equation}\label{lambda}
    \lambda = \rho_{22}\rho_{33}+|\rho_{23}|^2 \mp
    2\sqrt{\rho_{22}\rho_{33}}|\rho_{23}|~.
\end{equation}
    Thus, the four possible values of the $\lambda$ coefficients are
    as follows:
\begin{equation}\label{lambda4}
    \lambda_a = (\sqrt{\rho_{22}\rho_{33}}-|\rho_{23}|)^2, \quad
    \lambda_b = (\sqrt{\rho_{22}\rho_{33}}+|\rho_{23}|)^2, \quad
    \lambda_c = \rho_{11}\rho_{44}, \quad
    \lambda_d = \rho_{11}\rho_{44}.
\end{equation}
    Finally, the square roots of these lambda coefficients are directly employed in
    \eqref{eq:wottersformula2}.
    %...............
%\begin{equation}\label{sqroot}
%    \sqrt{\lambda_a} = \sqrt{\rho_{22}\rho_{33}}-|\rho_{23}|~,
%    \,\,
%    \sqrt{\lambda_b} = \sqrt{\rho_{22}\rho_{33}}+|\rho_{23}|~,
%    \,\,
%    \sqrt{\lambda_c} = \sqrt{\rho_{11}\rho_{44}}~, \,\,
%    \sqrt{\lambda_d} = \sqrt{\rho_{11}\rho_{44}}~.
%\end{equation}

    %\textcolor{red}{ahora solo falta poner la formula final explicando de manera convincente el orden de la resta. Para eso hay que analizar los operadores en forma de promedios, que tambien faltan}
    Noticing that $\lambda_b$ is the largest eigenvalue, the latter results leads immediately to Wootters' formula Eq.~(2).
%\begin{align}
 %   C &= \max \{ 0, \sqrt{\rho_{22}\rho_{33}}+|\rho_{23}| - \sqrt{\rho_{22}\rho_{33}}+|\rho_{23}| - \sqrt{\rho_{11}\rho_{44}} -
  %  \sqrt{\rho_{11}\rho_{44}} \} \\
  %  C &= \max \{ 0, 2|\rho_{23}| - 2\sqrt{\rho_{11}\rho_{44}} \} \\
   % C &= 2\max \{ 0, |\rho_{23}| - \sqrt{\rho_{11}\rho_{44}} \}
%\end{align}

%\begin{equation}\label{final_conc}
%C = \max \{ 0, 2|\rho_{23}| - 2\sqrt{\rho_{11}\rho_{44}}\}=2\max \{
%0, |\rho_{23}| - \sqrt{\rho_{11}\rho_{44}}\}
%\end{equation}

\clearpage

%fig1

\newpage

%fig2

\newpage

%fig3

\newpage

%fig4

\newpage

%fig5
\begin{figure}[htb]
\begin{center}
%\resizebox{135mm}{!}{\includegraphics{Final1-044-1-1-500-350-70-002-llenado-completo-v10.eps}}
\caption{(Color online) Best and average fitness for each generation
in a one-dimensional 44-site lattice subjected to periodic boundary
conditions, nearest-neighbor interactions, for $500$ generations,
population size of $350$.}  %$p_c=0.70$; $p_m=0.002$.}
\label{fig:Final1-044-1-1-500-350-completo}
\end{center}
\end{figure}

%\newpage

%fig5b - osea 6
\begin{figure}[htb]
\begin{center}
\resizebox{135mm}{!}{\includegraphics{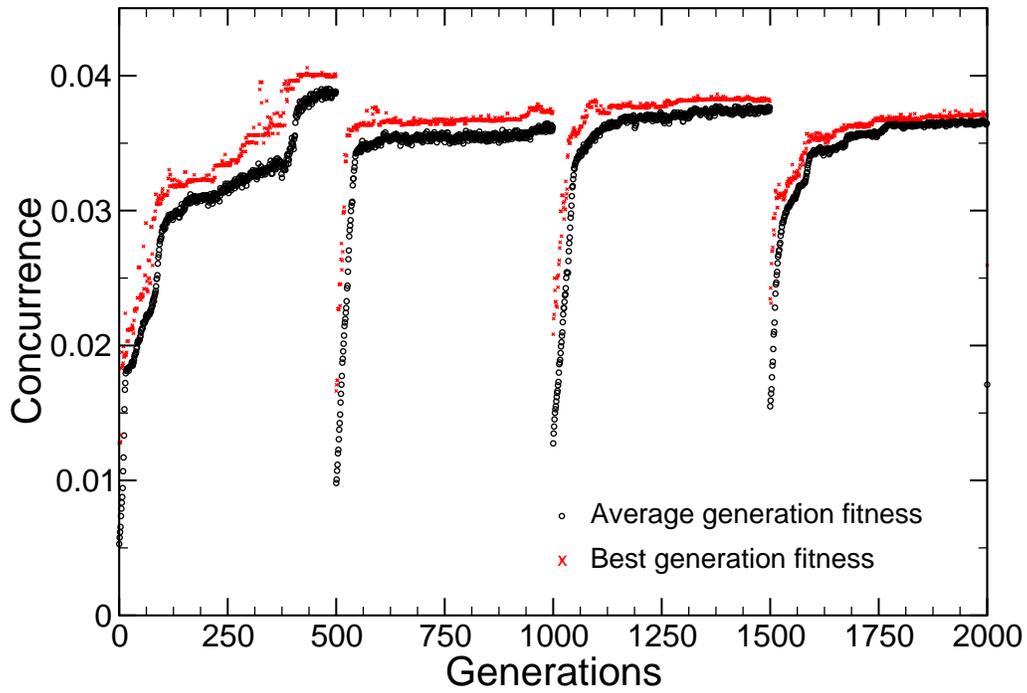}}
\caption{(Color online) Detail of
Fig.~\ref{fig:Final1-044-1-1-500-350-completo} for the first $2000$
generations.}
% Best and average fitness for each
%generation in a one dimensional 44 site lattice.}
\label{fig:Detalle1-044-1-1-500-350-completo}
\end{center}
\end{figure}

\newpage

%fig5
%\begin{figure}[htb]
%\begin{center}
%\resizebox{125mm}{!}{\includegraphics{FFF1-044-1-1-500-350-70-002-llenado-completo-v10.eps}}
%\caption{Best and average fitness for each generation in a one
%dimensional 44 site lattice.} \label{fig:1-044-1-1-200-400-completo}
%\end{center}
%\end{figure}

%fig7

\newpage

%fig8

\newpage

%fig9

\newpage

%fig10

\newpage

%fig11

\newpage

%fig12

\newpage

%fig6 osea fig. 13

\newpage

\end{document}